\pdfoutput=1
\documentclass[twocolumn,prd,floatfix,nofootinbib,longbibliography,notitlepage]{revtex4-1}

\usepackage{amsmath}
\usepackage{amssymb}
\usepackage{bm}
\usepackage{braket}
\usepackage{booktabs}
\usepackage{subfigure}
\usepackage{subfloat}
\usepackage{float}
\usepackage[usenames,svgnames,dvipsnames]{xcolor}

\usepackage{siunitx}
\usepackage{dsfont}

\usepackage{dcolumn}
\usepackage{hyphenat}

\usepackage{graphicx}

\graphicspath{{./Figures/}}

\usepackage[T1]{fontenc}
\usepackage[utf8]{inputenc}
\usepackage{lmodern}

\usepackage[unicode, colorlinks, allcolors=blue!70!black, linktocpage, pdfusetitle]{hyperref}
\hypersetup{
    colorlinks=true,
    urlcolor=SteelBlue,
    linkcolor=red,
    citecolor=blue,
}
\usepackage[all]{hypcap}

\usepackage{orcidlink}

\usepackage{microtype}

\newcommand{\Nsamp}{N_{\text{samp}}}
\newcommand{\DKL}{D_{\text{KL}}}
\newcommand{\avg}[1]{\langle#1\rangle}

\newcommand{\figschemesetup}{%
\begin{figure*}[t!]
\centering
\subfigure{
\includegraphics[width=.525\textwidth]{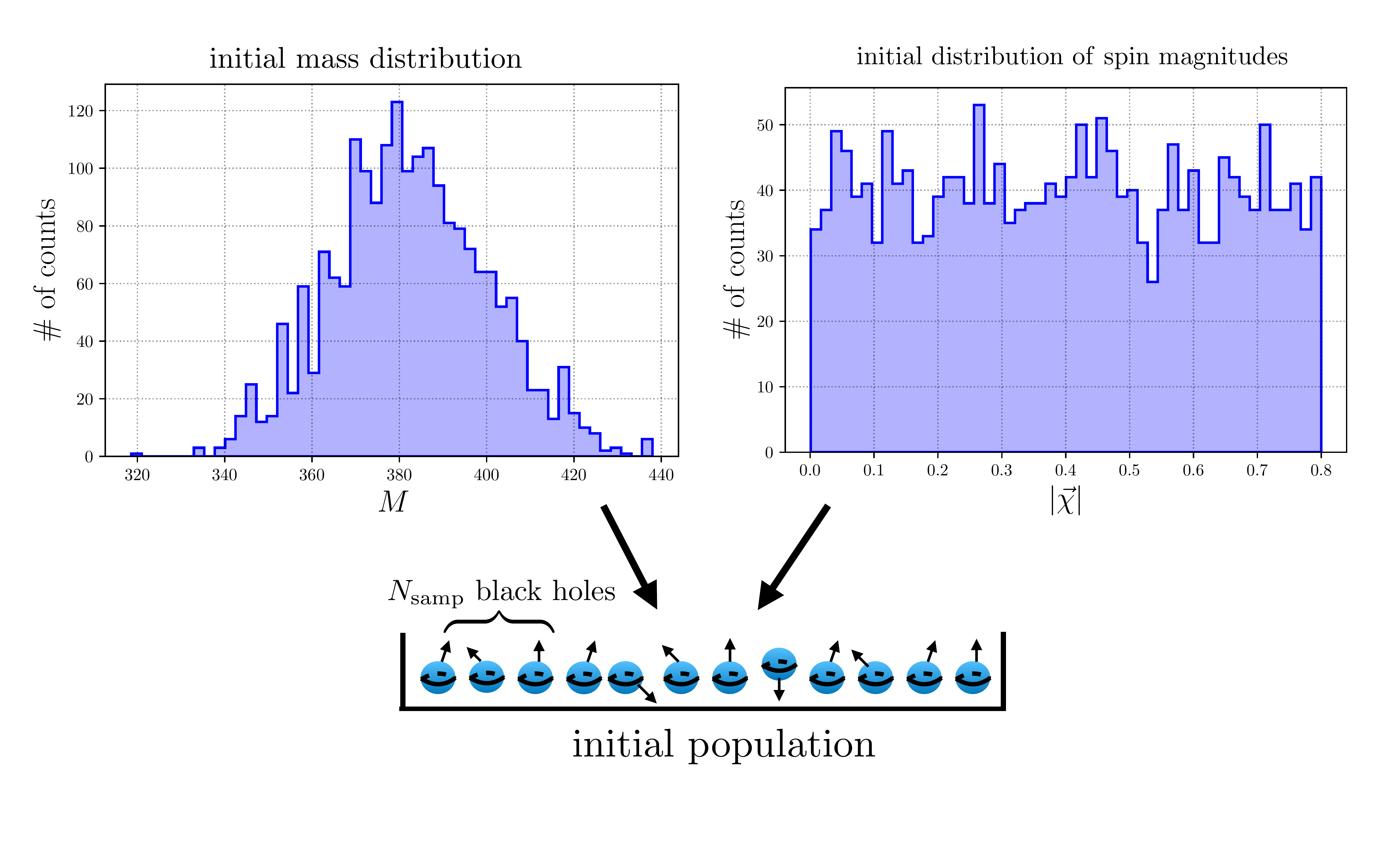}} \,
\subfigure{
\includegraphics[width=.365\textwidth]{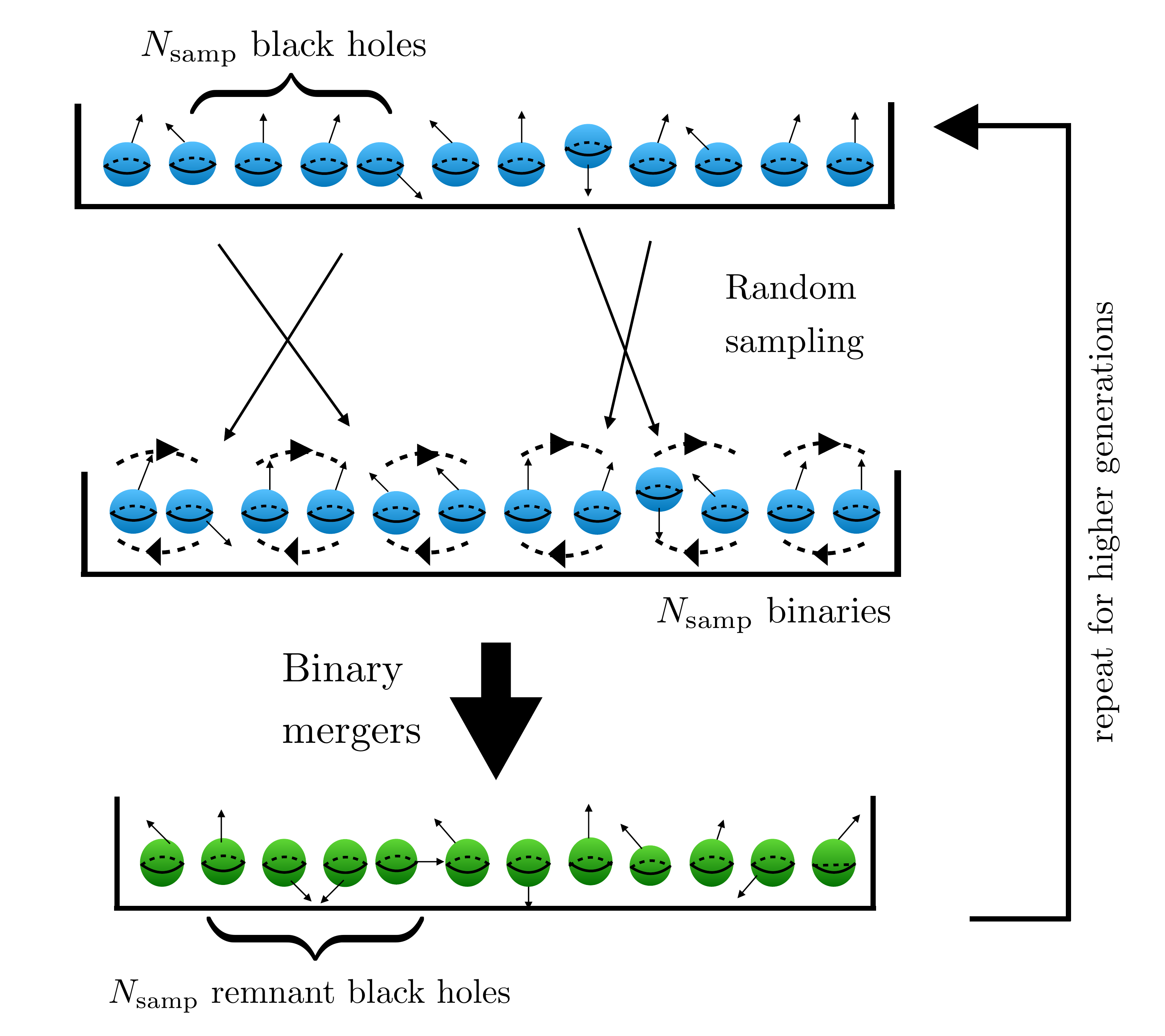}}
\caption{\label{fig:scheme_setup} Left: Preparation of initial
  conditions from distributions of spins and masses.  We assume an
  isotropic collection of black holes, meaning the spins are
  distributed uniformly in solid angle.  Right: Resampling for
  pairs of black holes to merge, the outcome giving a subsequent
  generation of black holes.  From a finite sample of BHs, we
  can randomly form $\Nsamp$ binaries (each BH chosen independently)
  to generate another $\Nsamp$ remnant BHs for the next generation. The
  ``box'' prepared in the left panel is the same as the box on the top of the 
  right panel, providing initial conditions for the first generation
  of mergers.
  \vspace{-1em}
}
\end{figure*}
}

\newcommand{\figevolhistmassspin}{%
\begin{figure*}[t!]
\centering
\subfigure{
\includegraphics[width=.415\textwidth]{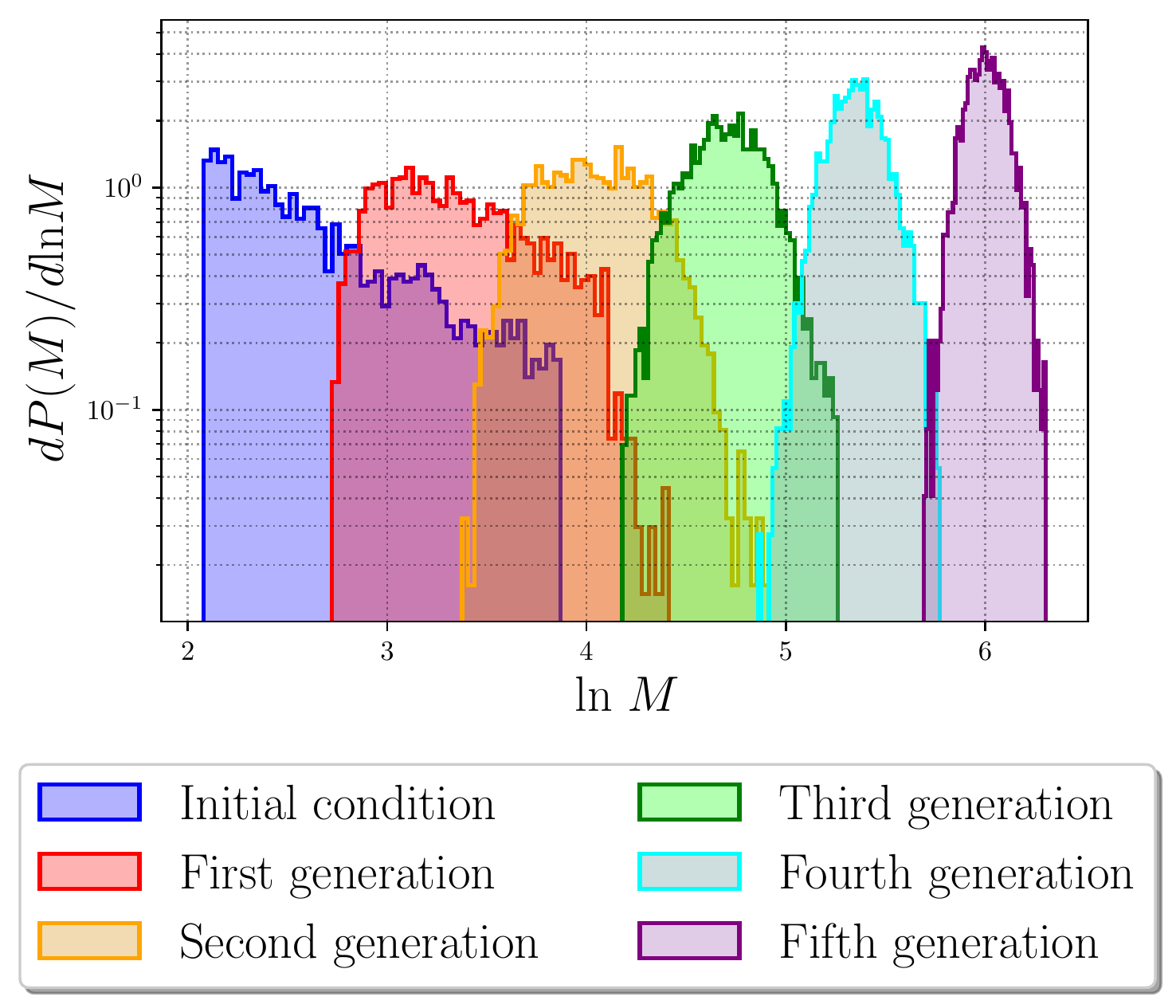}} \,
\subfigure{
\includegraphics[width=.435\textwidth]{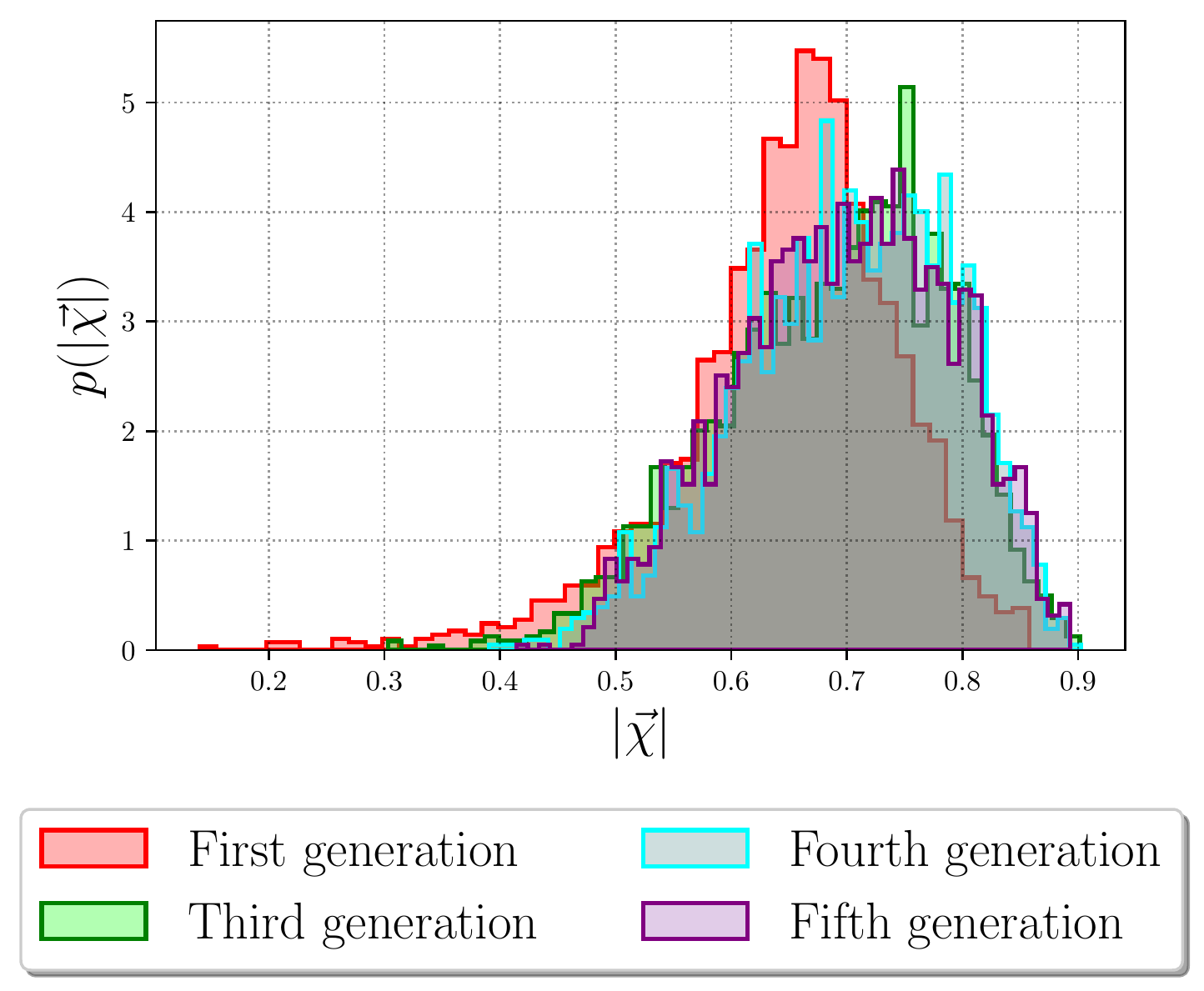}}
\caption{\label{fig:evol_hist_mass_spin}%
  Evolution of the marginalized mass (left) and spin (right)
  distributions with generation.  The initial conditions were a Kroupa
  mass distribution, and uniform in the magnitude $\chi$.  The mass
  distribution gets narrower (in log space) with the number of
  generations.  There is no noticeable difference in marginalized spin
  distribution from the third to the fifth generation, consistent with
  approaching the attractor distributions.
  \vspace{-0.5em}
}
\end{figure*}
}

\newcommand{\figPDFmuchi}{%
\begin{figure*}[t!]
\centering
\includegraphics[width=.77\textwidth]{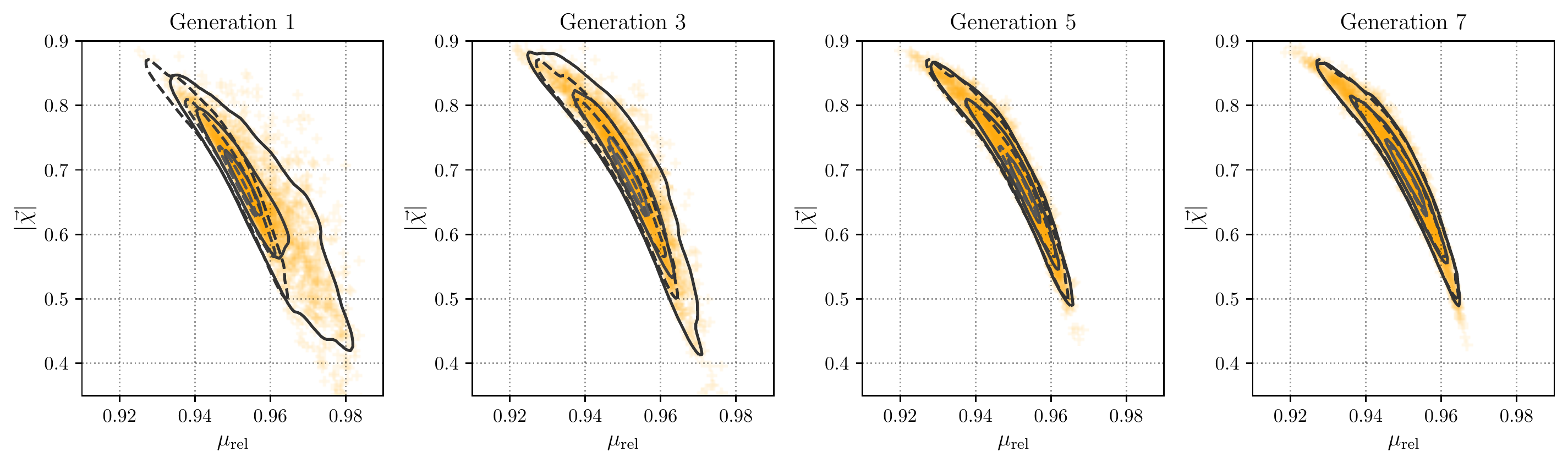}
\caption{Evolution of the joint distribution
  $p(\mu_{\mathrm{rel}},\chi)$.  The solid contours enclose 30\%, 68\%,
  and 95\% of the probability mass of the $n$th generation.  Dashed
  contours are from the fixed-point distribution.  Each cross in the
  orange diffuse cloud represents a remnant black hole.%
  \vspace{-0.5em}
\label{fig:PDF_mu_chi}%
}
\end{figure*}
}

\newcommand{\figevolfixedmassspin}{%
\begin{figure*}[t!]
\centering
\subfigure{
\includegraphics[width=.435\textwidth]{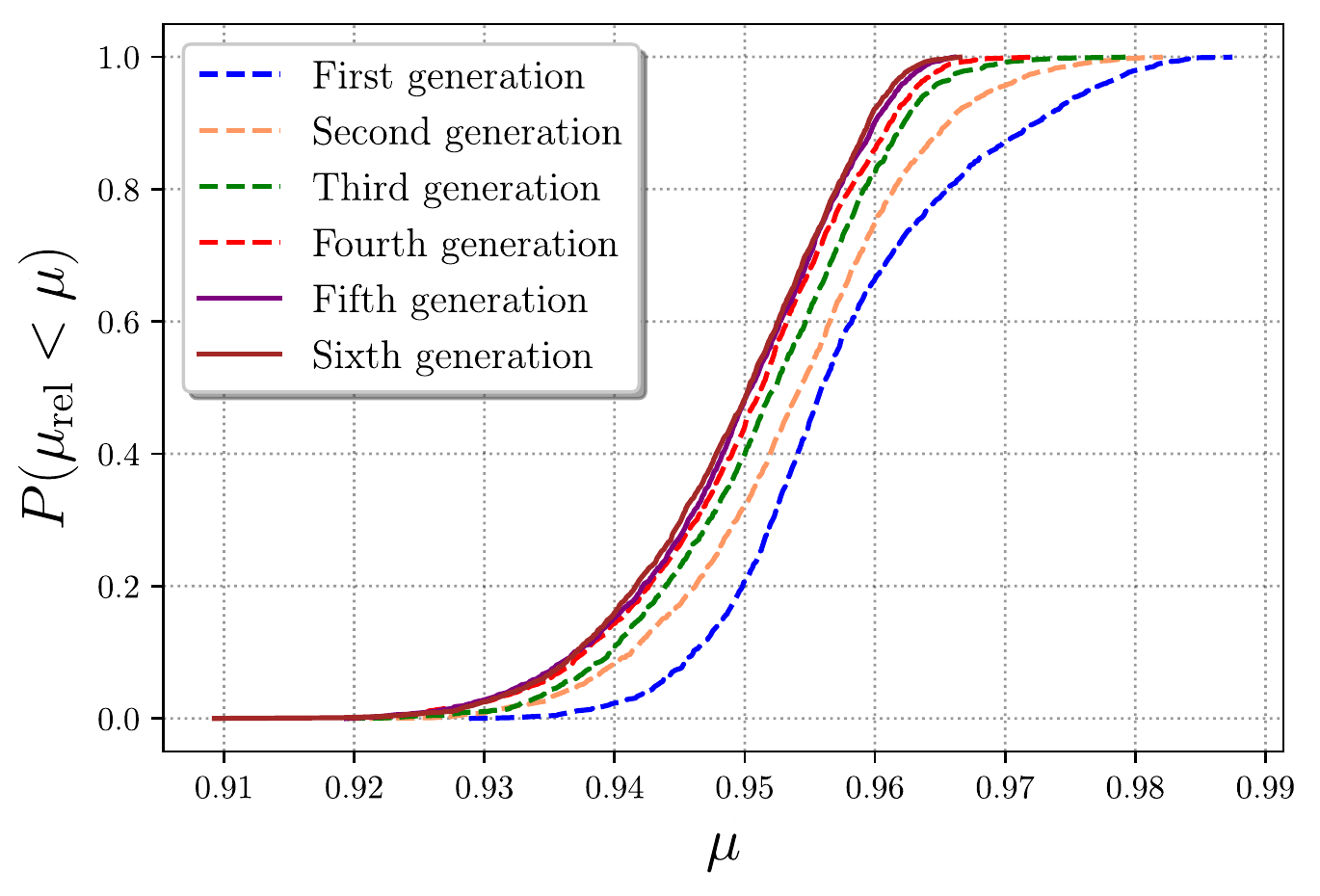}} \,
\subfigure{
\includegraphics[width=.43\textwidth]{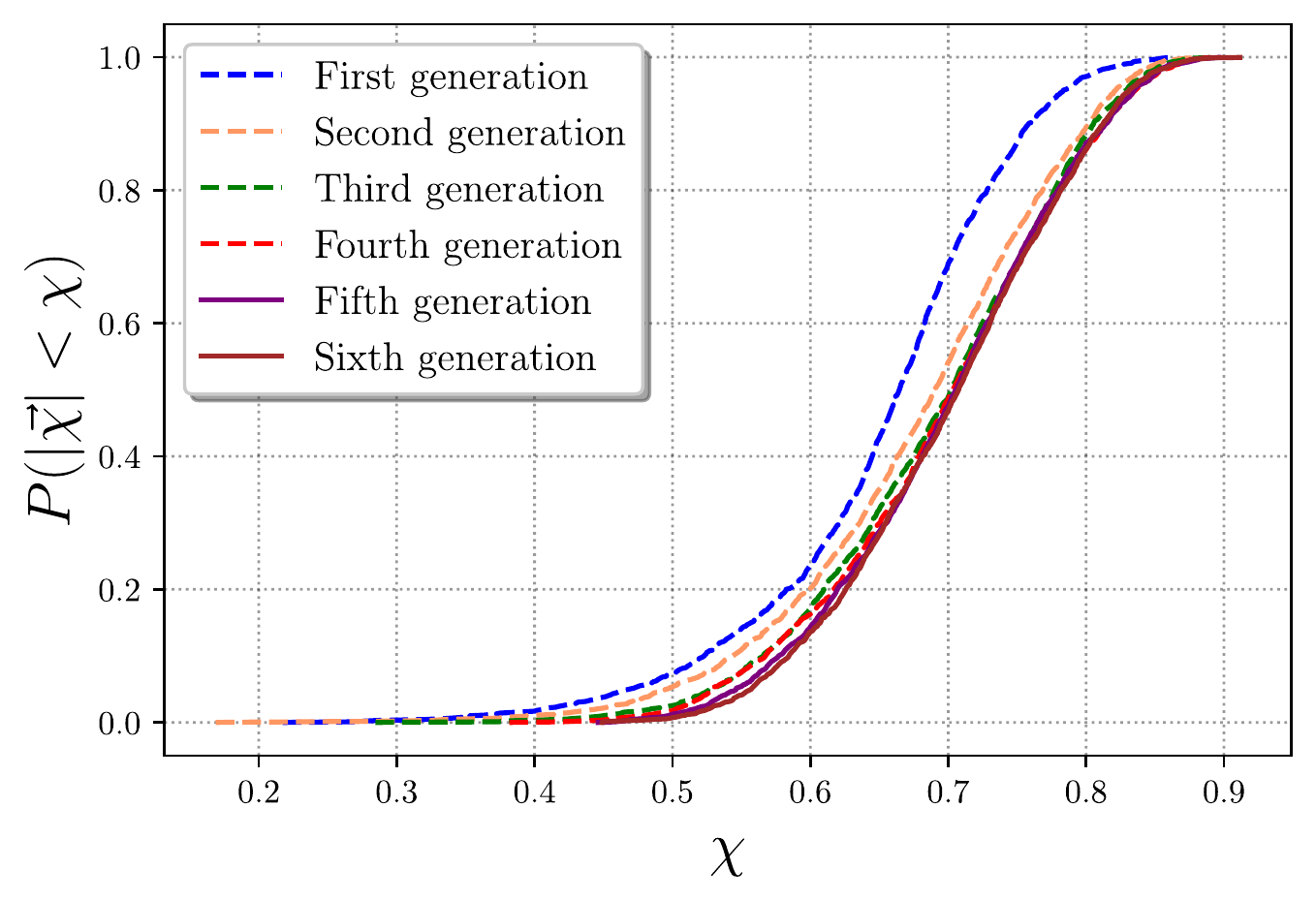}}
\caption{\label{fig:evol_fixed_mass_spin}%
  Evolution of the cumulative relative mass ratio distribution
  $P(\mu_{\mathrm{rel}}<\mu)$ (left) and spin magnitude distribution
  $P(|\vec{\chi}|<\chi)$ (right) as a function of number of
  generations.  The initial mass distribution was the Kroupa power
  law, for which it takes six generations to converge to the
  fixed-point.  We plot the first four generations with dashed lines.
  \vspace{-1em}
}
\end{figure*}
}

\newcommand{\figKLmassspin}{%
\begin{figure*}[t!]
\centering
\subfigure{
\includegraphics[width=.48\textwidth]{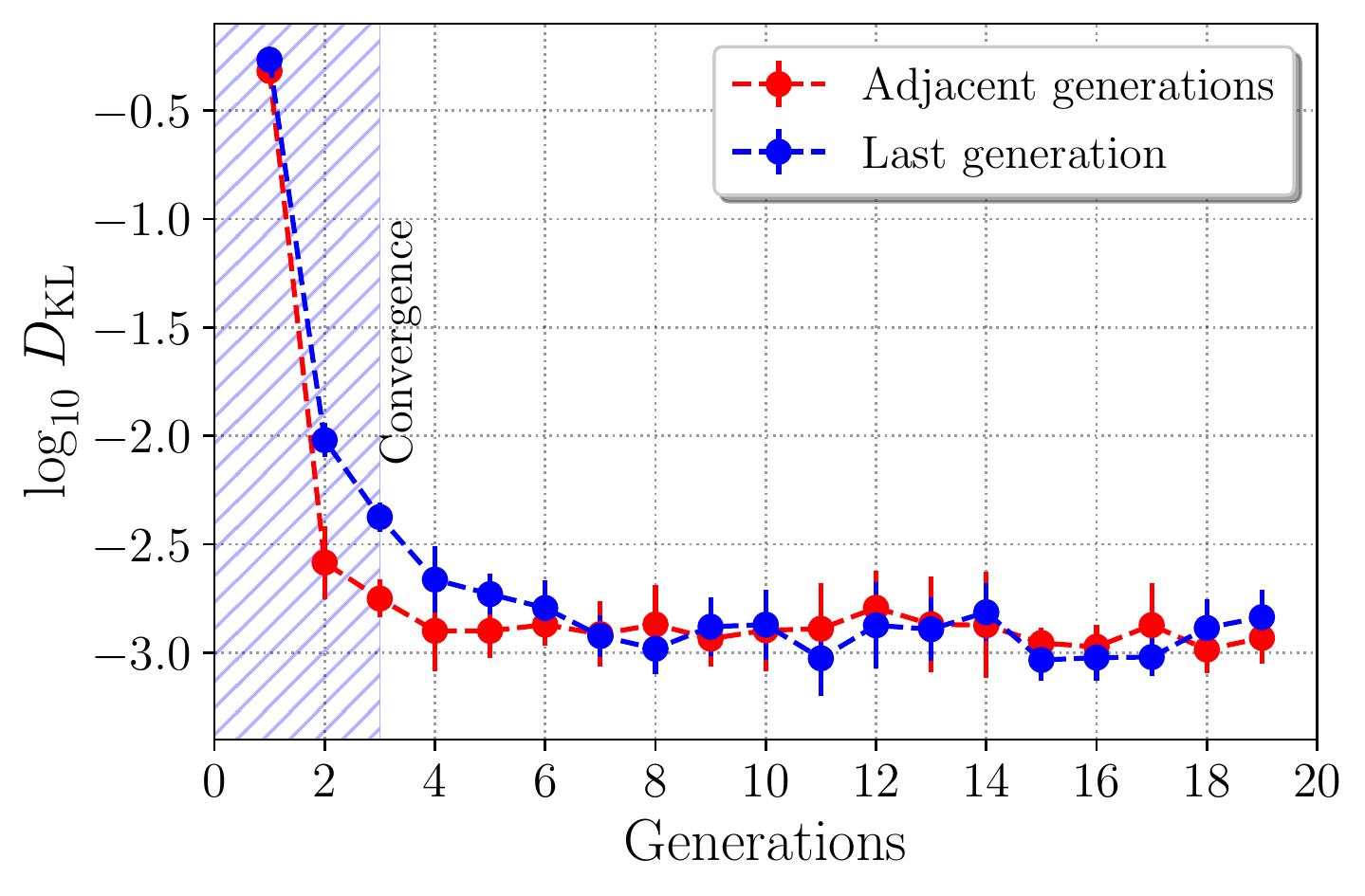}} \,
\subfigure{
\includegraphics[width=.47\textwidth]{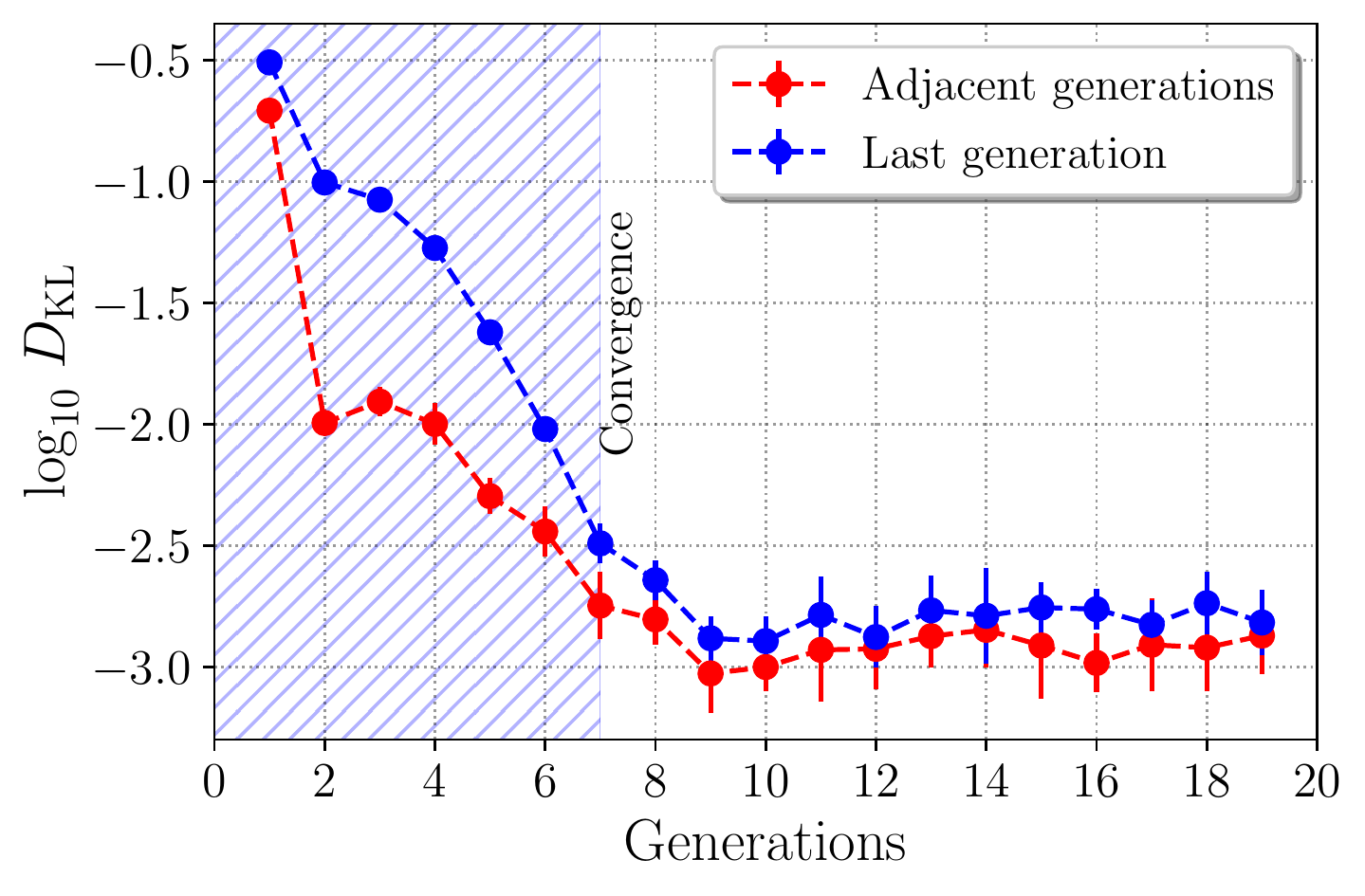}}
\caption{\label{fig:KL_mass_spin}%
  Values of $\DKL$ between 2D joint PDFs
  from adjacent generations, $\DKL(i+1||i)$, (red); or the PDF of
  generation $i$ relative
  to the last (19th) generation, $\DKL(19||i)$, (blue).  Each dashed
  region contains the points needed to converge to a joint 
  fixed-point distribution. Error bars are estimates of KDE
  finite sampling effects, estimated by subsampling 50\% of the
  $\Nsamp=40,000$ binaries in each generation several times, and
  computing the variance.
  Left: Initial conditions are a
  Gaussian in mass.  It only takes two
  generations to converge to a fixed-point distribution.
  Right: Initial conditions are a Kroupa power-law in mass.  Here it
  takes at least seven or eight generations to converge to the same
  fixed-point distribution.
\vspace{-1em}
}
\end{figure*}
}

\newcommand{\figconvinitcond}{%
\begin{figure}[t!]
\centering
\includegraphics[width=.43\textwidth]{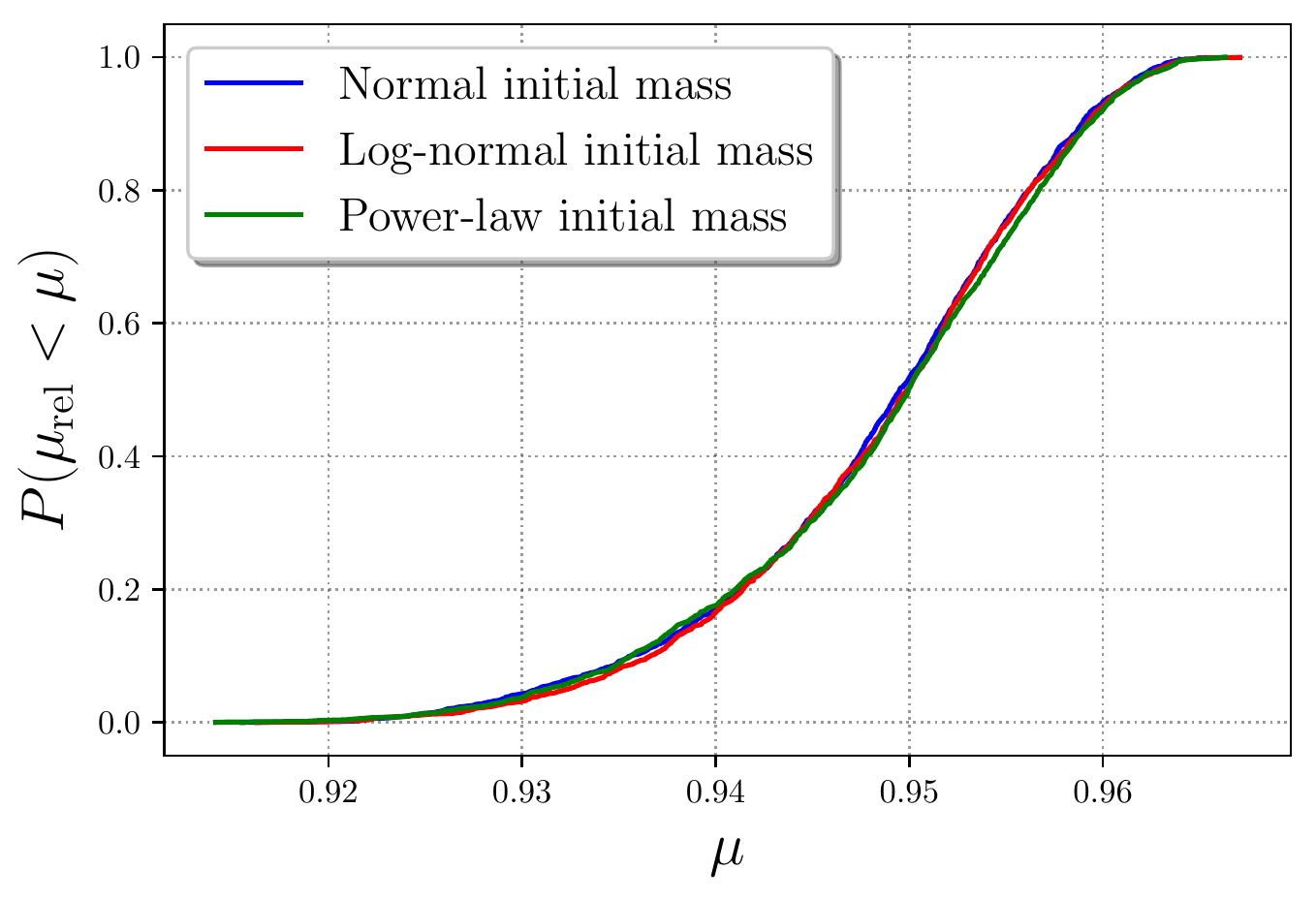}
\caption{The cumulative distribution
  function of $\mu_{\mathrm{rel}}$ is independent of initial
  conditions. The same independence applies to the spin
  magnitude and the kick velocity distributions.
  \label{fig:conv_init_cond}
  \vspace{-0.75em}
}
\end{figure}
}

\newcommand{\figlogevol}{%
\begin{figure}[t!]
\centering
\includegraphics[width=.435\textwidth]{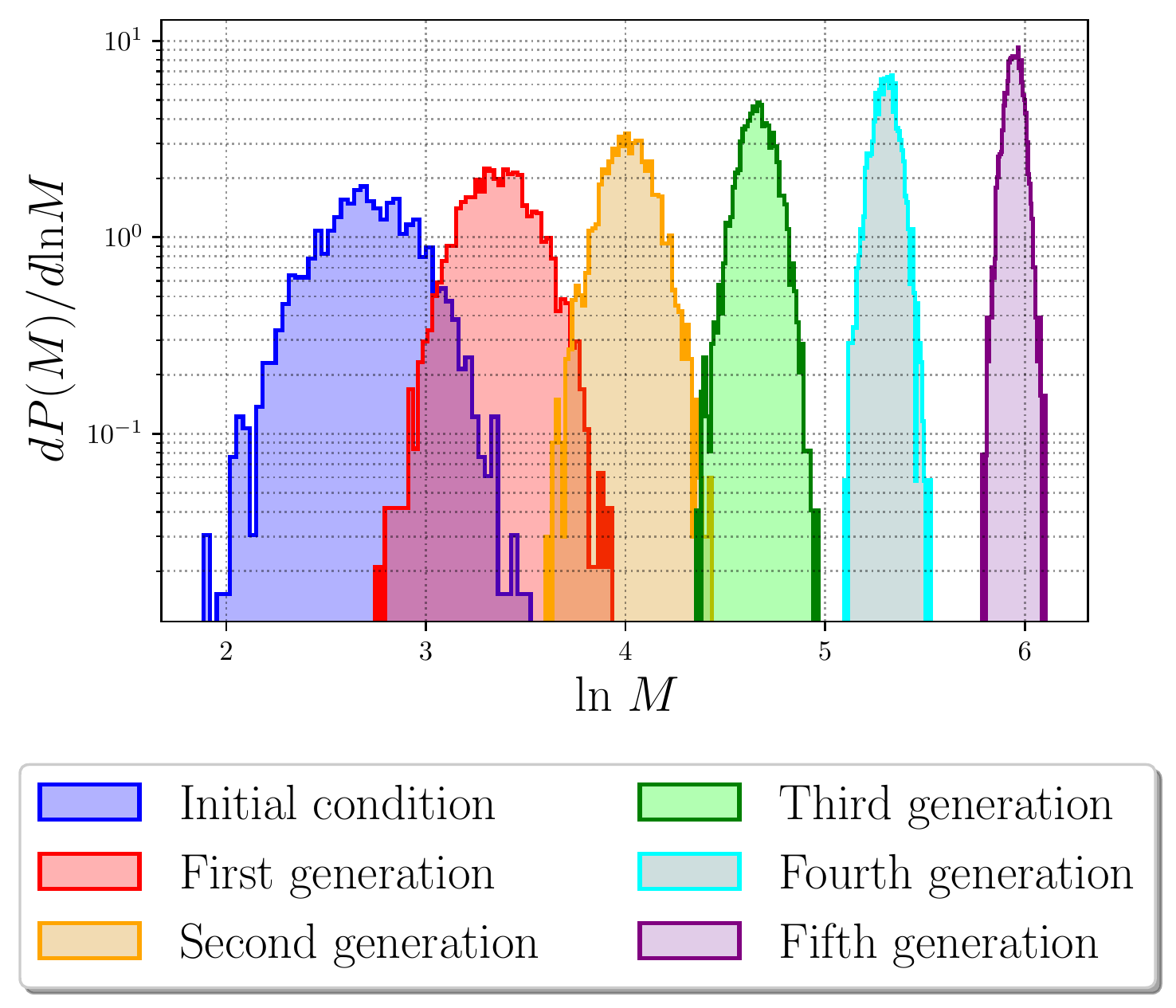}
\caption{Evolution of mass distribution, with a log-normal initial
  condition.  Each generation's shape is consistent with
  log-normal.  The evolution from an initial Gaussian is
  similar.
  \label{fig:log_evol}
  \vspace{-1.5em}
}
\end{figure}
}

\newcommand{\figthermal}{%
\begin{figure*}[t!]
\centering
\subfigure{
\includegraphics[width=.455\textwidth]{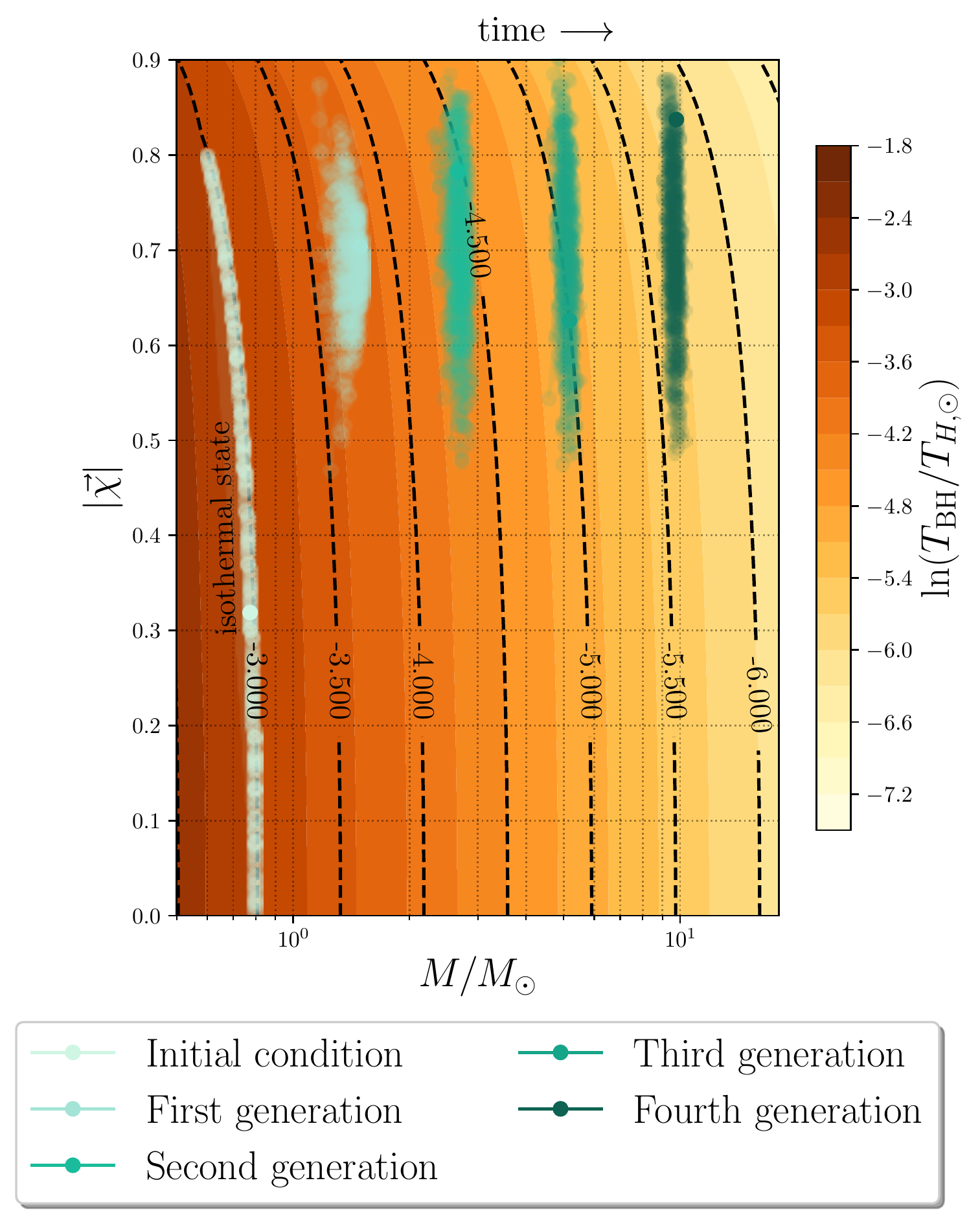}} \,
\subfigure{
\includegraphics[width=.457\textwidth]{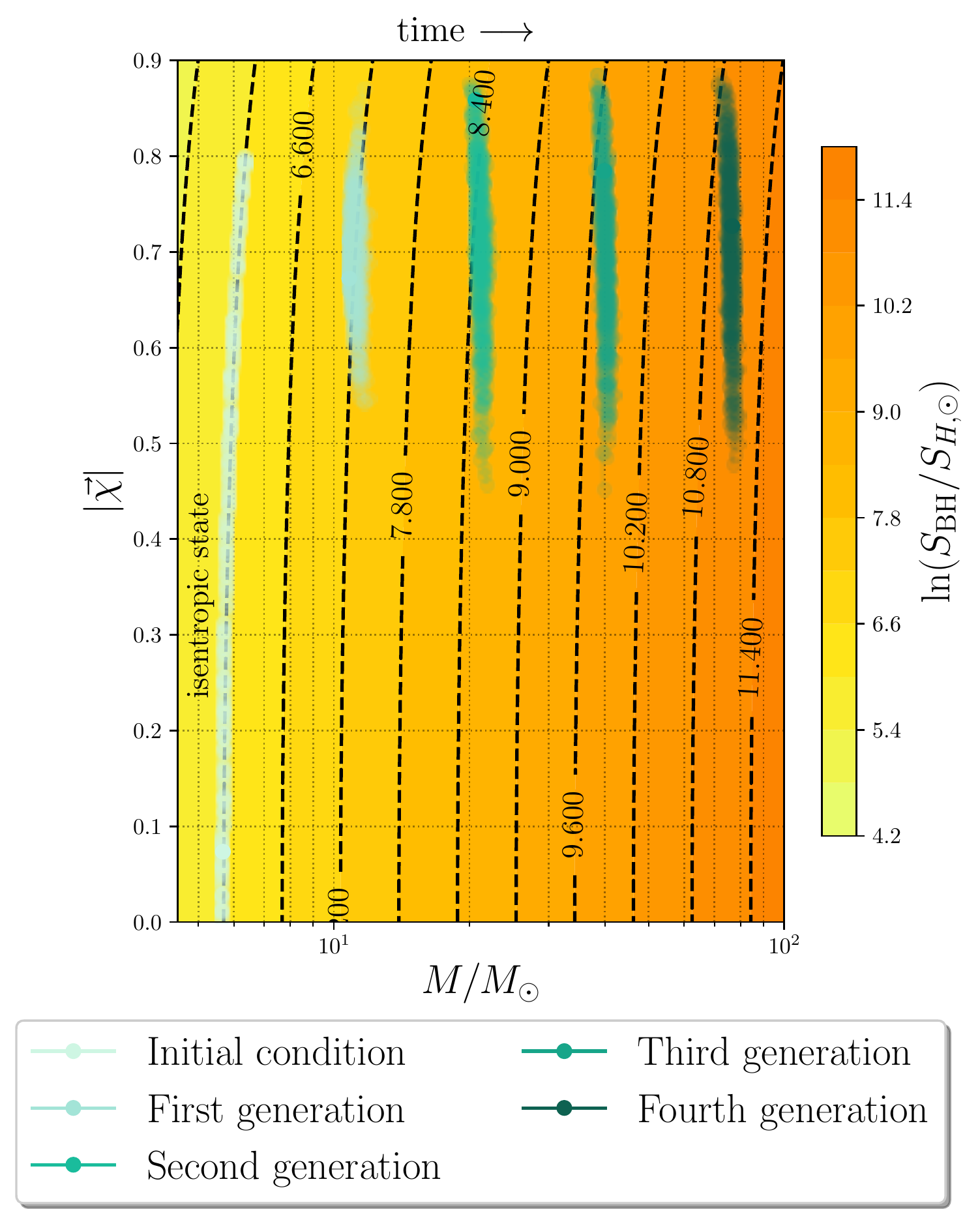}}
\caption{\label{fig:thermal}%
  Left: Evolution of an isothermal collection of black holes. Right:
  Evolution of an isentropic collection of black holes. All the
  elements in the initial configuration belong to a constant
  temperature/entropy curve. In later generations, irreversible
  non-linear dynamics dilutes the thermal properties of the
  configurations, no longer on isothermal/isentropic curves, and they
  approach the fixed point distribution.
\vspace{-1em}
}
\end{figure*}
}

\newcommand{\figentropy}{%
\begin{figure}[t!]
\centering
\includegraphics[width=.45\textwidth]{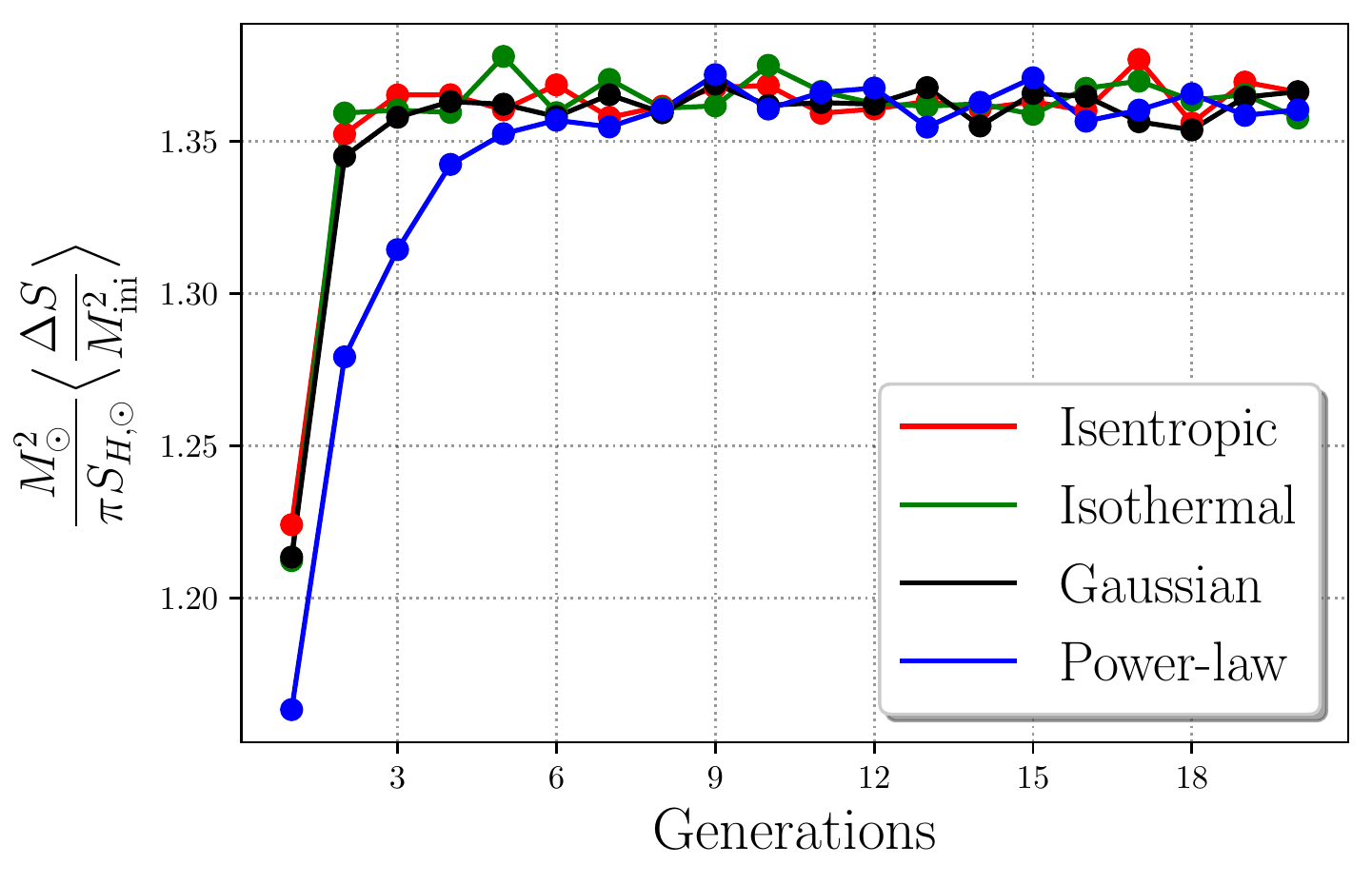}
\caption{\label{fig:ent_evol}Evolution of the entropy production rate
  per merger for different choices of initial distributions. This
  quantity converges more slowly when the initial mass distribution is
  the Kroupa power law.
\vspace{-1em}
}
\end{figure}
}

\newcommand{\figfulljointPDF}{%
\begin{figure}[t!]
\centering
\includegraphics[width=.5\textwidth]{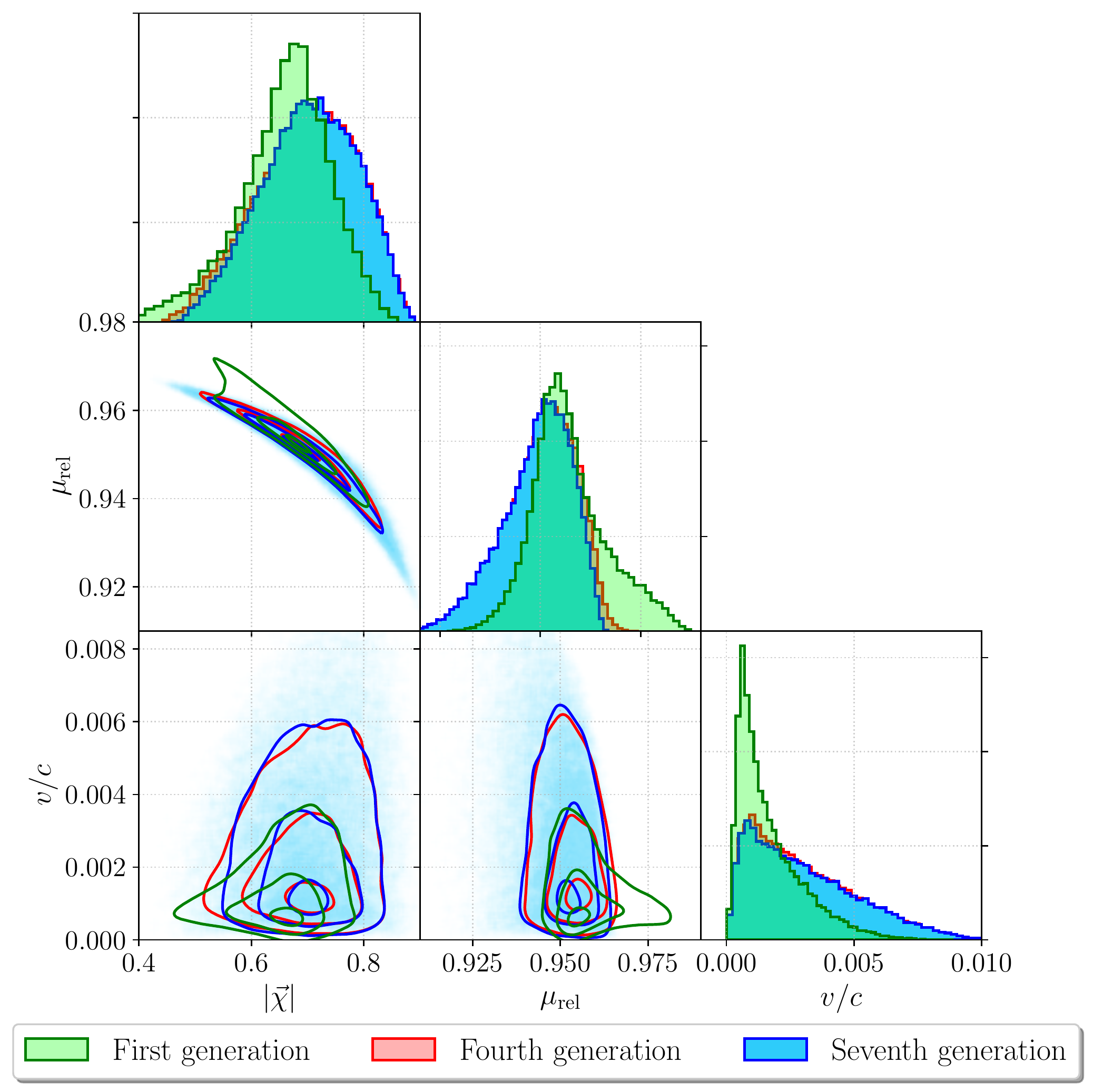}
\caption{\label{fig:corner_mu_chi_v}Corner plot showing the joint distribution 
$p(\mu_{\mathrm{rel}},\chi, v/c)$ after one, four and seven generations of 
mergers. Nested contours (in green for the first, in red for the fourth and 
in blue for the seventh generation) contain 30\%, 68\% and 95\% of the total 
probability mass. Each point in the light blue cloud represents a black hole 
remnant drawn from the fixed-point distribution.}
\end{figure}
}

\newcommand{\figmoments}{%
\begin{figure*}[t!]
\centering
\subfigure{
\includegraphics[width=.43\textwidth]{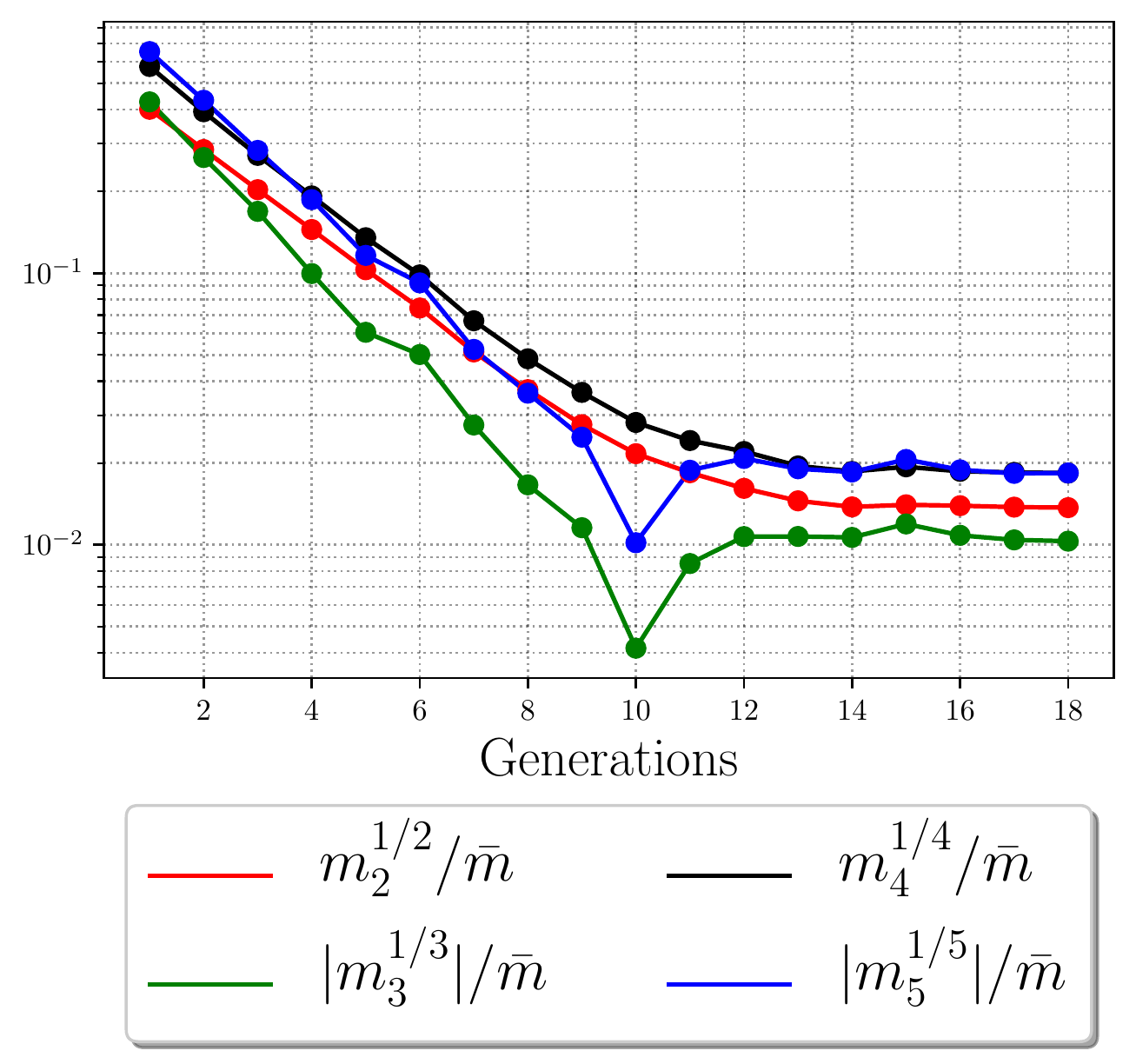}} \,
\subfigure{
\includegraphics[width=.448\textwidth]{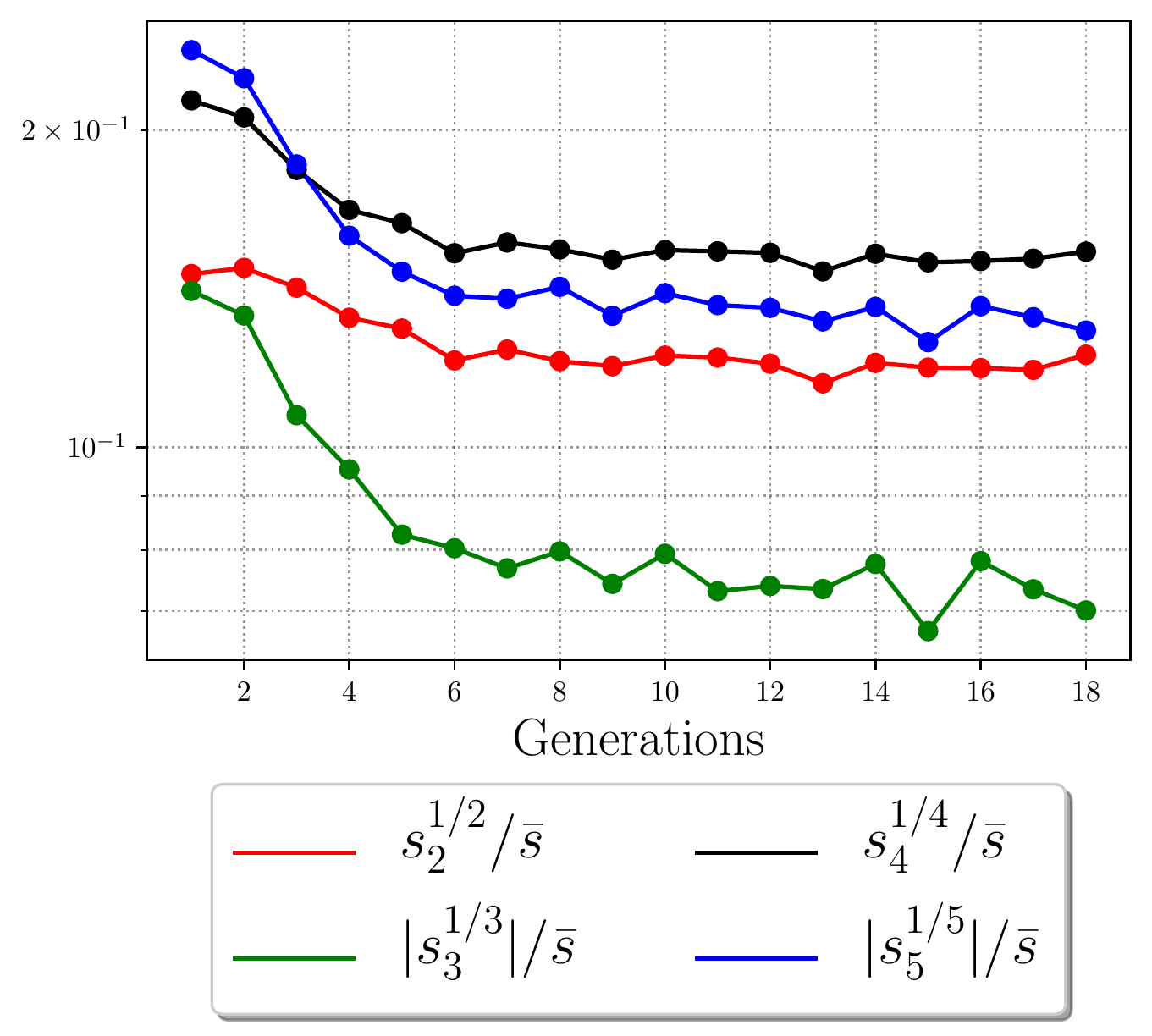}}
\caption{\label{fig:moments}%
Central mass (left) and spin (right) moments, made dimensionless in
units of mean mass and spin, as a function of generation.
The mass moments all seem to decrease until the twelfth generation, where the moments
become constant.  The spin moments converge more quickly.
\vspace{-1em}
}
\end{figure*}
}

\newcommand{\figvindep}{%
\begin{figure}[t!]
\centering
\includegraphics[width=.48\textwidth]{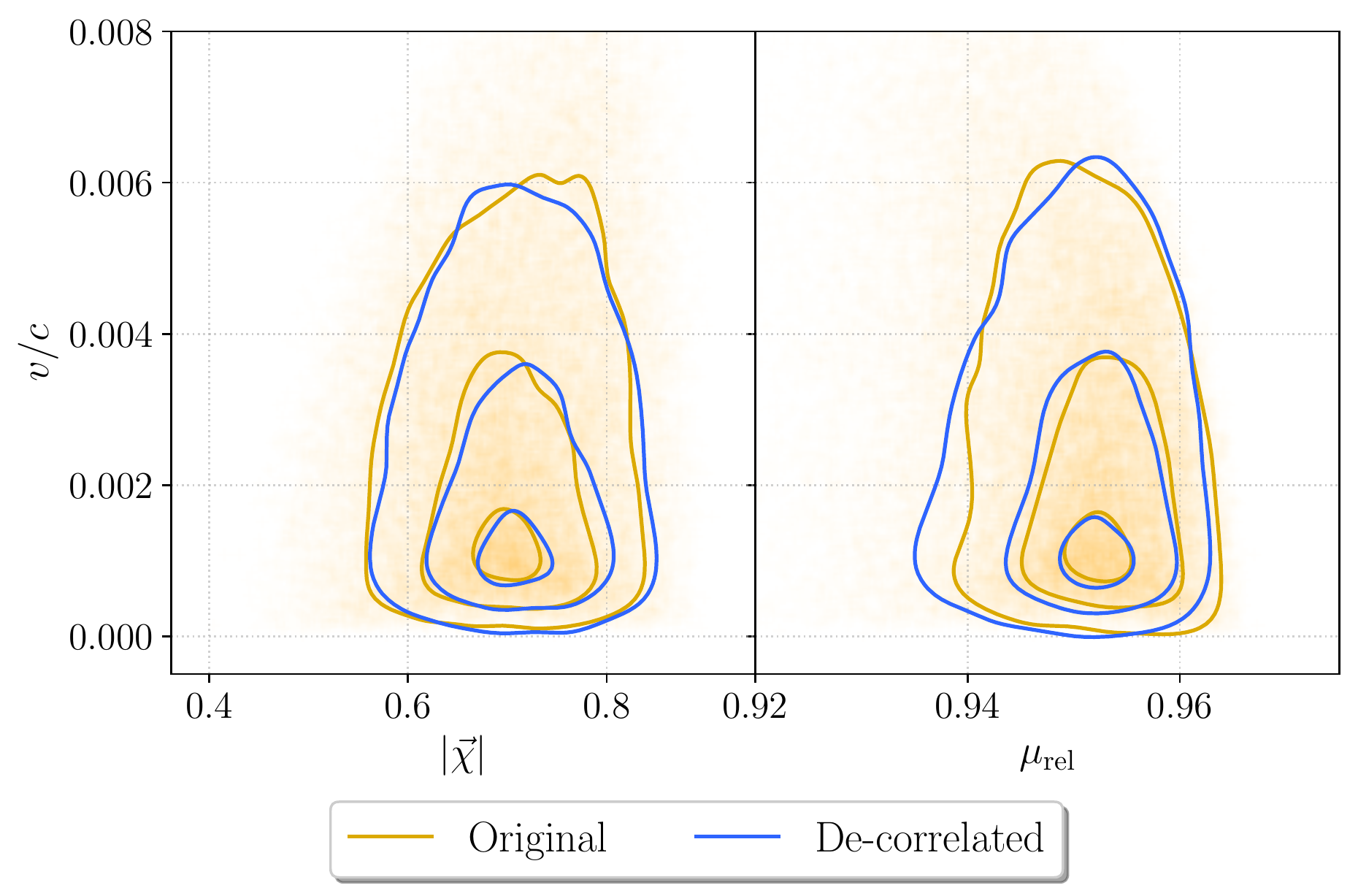}
\caption{Comparing the joint distributions $p(\chi, v/c)$ and $p(\mu_{\mathrm{rel}}, v/c)$ from
the fixed-point distribution with a sample with kick velocities decorrelated from the
other parameters. The differences in the nested contours of the original 
marginalized distributions (in orange) and the contours from the samples with de-correlated
velocities (in blue) are small, consistent with an approximate independence of
the kick velocity. Small correlations are only visible in the tails in the right panel.
  \label{fig:v_indep}
  \vspace{-1.5em}
}
\end{figure}
}

\begin{document}

\title{A fixed point for black hole distributions}

\author{Jos\'e T. G\'alvez Ghersi\,\orcidlink{0000-0001-7289-3846}}
\email{jgalvezg@phy.olemiss.edu}
\author{Leo C. Stein\,\orcidlink{0000-0001-7559-9597}}
\email{lcstein@olemiss.edu}
\affiliation{Department of Physics and Astronomy, The University of Mississippi,
University, MS 38677, USA}
\date{\today}

\hypersetup{pdfauthor={G\'alvez and Stein}}

\begin{abstract}
Understanding distributions of black holes is crucial to both
astrophysics and quantum gravity.  Studying astrophysical population
statistics has even been suggested as a channel to constrain
black hole formation from the quantum vacuum.
Here we propose a Gedankenexperiment to show that the non-linear
properties of binary mergers (simulated with accurate
surrogate models) generate an attractor in the space of
distributions.
Our results show that the joint distribution of
spin magnitude and fractional mass loss evolves to a fixed point,
converging in a few generations. The features of this
fixed point distribution do not depend on the choice of initial
distribution.
Since a black hole merger is irreversible it produces entropy --
possibly the largest source of entropy in the universe.
The fixed-point distributions are neither isothermal nor isentropic, and
initially thermodynamic states evolve away from thermality.
We finally evaluate the specific entropy production rate per merger
from initially thermal and non-thermal distributions, which converges
to a constant.
\end{abstract}
\maketitle

\section{Introduction}
Black hole distribution functions carry a wealth of information, both
in the theoretical context of quantum gravity, and the observational
context of astrophysics.  Since the work of Bekenstein and
Hawking~\cite{Bekenstein:1973ur, Bekenstein:1974ax, Bekenstein:1975tw,
  Hawking:1974sw}, the interpretation of black holes as thermodynamic
and statistical mechanical systems has established black holes as a
key to understanding quantum gravity.  Their thermal properties have
lent weight to top-down calculations of the microscopic degrees of
freedom of quantum gravity~\cite{Strominger:1996sh, Ashtekar:1997yu}.
Meanwhile, in an astrophysical context, BH distribution functions
teach us about stellar populations, the environments in which stars
live, die, and give birth to BHs; dynamical interactions, and
more~\cite{Fryer:1999ht, Berti:2008af, Gerosa:2018wbw,
  Kimball:2019mfs, Kimball:2020opk, Gerosa:2019zmo, Rodriguez:2019huv,
  Banerjee:2020bwk}.  The possibility of primordial BHs (PBHs) lie at
the intersection of quantum gravity and astrophysics.  If they exist,
their population may be an important component of dark matter, a
question which can now be addressed with observations from Advanced
LIGO and Virgo~\cite{Carr:2020xqk}.  The quantum-mechanical and
statistical properties of PBHs may leave imprints on astrophysical
observations~\cite{Raccanelli:2017xee}.  Conversely, observations
could test whether PBHs were seeded by a quantum thermodynamic
process, producing PBHs in a (micro)canonical
ensemble~\cite{Bianchi:2018ula}.

In this paper, we present a fundamental black hole distribution
function, determined solely by the non-linear, irreversible,
out-of-equilibrium process of the merger of two black holes.
Binary black hole (BBH) mergers are the dominant source
in gravitational-wave astronomy~\cite{Abbott:2016blz, Abbott:2017vtc, TheLIGOScientific:2017qsa,
  LIGOScientific:2018mvr, LIGOScientific:2018jsj,
  LIGOScientific:2020stg},
which enables the observational study of black hole distributions.
Numerical relativity (NR) simulations~\cite{Boyle:2019kee} are necessary for modeling
gravitational waveforms from binary black hole (BBH) mergers.
However, NR simulations require substantial computational
power and time to produce accurate results, which is an obstacle when
spanning the whole seven-dimensional parameter space of quasicircular
BBH mergers. Surrogate models~\cite{Varma:2018aht, Varma:2019csw}
interpolate NR simulation data to compute both the gravitational waveform and the properties 
of the black hole remnant
at points in parameter space even where no NR simulations were
performed. These models are efficient enough to evaluate populations
of thousands of BBH mergers in seconds
on laptop/desktop computers. 

We pioneer the use of surrogates to study black hole distribution
functions.
We simulate a thought experiment on a large population of black holes, where we choose initial
spin and mass distributions.  From this initial population, we configure a set of binaries by randomly
resampling for BH pairs, and then allowing each pair to merge. The only condition we require is
that the mass ratio of each binary is in the interval $1\leq q\leq 6$,
where the results of the \texttt{NRSur7dq4Remnant} surrogate
have been validated~\cite{Varma:2019csw}. We compute both the mass and the spin of each of the remnant black 
holes in the post-merger population using the surrogate model, rather than semiempirical
expressions or analytical/numerical kludges~\cite{Tichy:2008du, Barausse:2012qz,
  Hofmann:2016yih, Gerosa:2016sys,Campanelli:2007ew, Gonzalez:2006md,
  Lousto:2007db, Lousto:2012su, Lousto:2012gt}.
We then consider the remnant
parameter distributions as new initial conditions for the next
iteration to repeat this procedure, again resampling for binaries
merging them,
and find the mass and spin of every remnant in the next generation
of black holes. Hence it is possible to visualize how the spin and
mass distributions change with the number of generations.  Our results show 
the distributions of spin magnitudes and relative mass ratios converge
to fixed-point functions after a few generations.  We quantitatively test
for convergence using the Kullback-Leibler (KL) divergence~\cite{ref1}
in the 2D distributions, finding that (a)~our sampled distributions are
consistent with being drawn from the same underlying continuous
distribution; and (b)~different choices of initial mass distributions
only affect the number of generations needed to converge.

\figschemesetup

Some initial distribution functions are motivated probabilistically or astrophysically.
Meanwhile, the Bekenstein-Hawking (BH) formulae for entropy and
temperature~\cite{Bekenstein:1973ur, Hawking:1974sw} for the Kerr
solution allow us to define a population of black holes
at constant temperature (or entropy). 
These distributions define isothermal
(isentropic) curves in the mass-spin plane, similar to isothermal
(isentropic) curves in a PV phase
diagram for a gas in classical thermodynamics.  After computing the
mass and spin of remnant black holes for a few generations of binary
mergers, we find that resulting generations are not
constant temperature (entropy).
This is due to the irreversibility and out-of-equilibrium dynamics of
the highly non-linear merger process.
Knowing the mass and spin of each remnant black hole, we can compute the
entropy production rate of the distribution
in each generation.  Our results show that the specific entropy produced per
merger is constant after a certain
number of generations, depending on the shape of the initial mass
distribution.  This disfavors the presence of memory (in the Markov
sense, not the BMS sense) in late-generation mergers.

The layout of this paper is as follows: in Section~\ref{sec:setup}, we
describe in full detail the setup of our thought experiment.  In
Section~\ref{sec:fixed-point}, we use the setup described in the
previous section to show the existence of attractors in distribution
space.  We find that arbitrary distributions of mass and spin in a
large population of black holes converge to fixed-point distributions
of spin magnitude and relative mass ratios.  For consistency, we
use the Kullback-Leibler divergence to show that the distributions
require a few generations to converge.  The number of generations
depends on the shape of the initial mass distribution.  In
Section~\ref{sec:thermal}, we evaluate the evolution of samples of
black holes at constant Hawking temperature (entropy) forming an
isothermal (isentropic) state.  After successive binary mergers, we
observe that the out-of-equilibrium, non-linear nature of the merging
process suppresses the thermal features of such configurations.  We
also compute the entropy production rate per merger.  We discuss
and conclude in Section~\ref{sec:discussion}.

\section{The thought experiment}\label{sec:setup}
Here we describe the setup of our proposed thought experiment to study
the flow of distributions over merger generations.  The surrogate
model we use~\cite{Varma:2019csw} is built on a seven-dimensional
parameter space, consisting of two 3-D spin vectors and the mass ratio
$q$.  This model was trained on NR simulations in the range of
mass-ratio spanning $1\leq q \leq 4$, while faithfully extending up to
$q\leq 6$ by extrapolation.  For dimensionless spin magnitudes
$\chi = |\vec{\chi}|$, the model was built in the range
$0\leq \chi \leq 0.8$, and extrapolates reasonably up to
$\chi \leq 1.0$. As shown in~\cite{Varma:2019csw} (cf. Figs. 10 and 11),
the surrogate code is more precise than other validated fits such as \texttt{SEOBNRv3}.
Going to more extreme mass ratios would require
either semianalytical models~\cite{Hughes:2002ei, Barausse:2012qz, Hofmann:2016yih,
  Gerosa:2016sys,Campanelli:2007ew, Gonzalez:2006md, Lousto:2007db,
  Lousto:2012su, Lousto:2012gt} or a new surrogate built in the style
of~\cite{Rifat:2019ltp}.  However, such models have not been validated
 against NR simulations at higher mass-ratios, e.g.~\texttt{SEOBNRv3} has
 been validated up to $q=5$~\cite{Babak:2016tgq}.

The limitations of \texttt{NRSur7dq4Remnant} inform what
choices we can make for initial distributions.  We select an initial
population of black holes by assigning a mass $M$ and dimensionless
spin vector $\vec{\chi}$ to each, sampled from a given distribution
function, as shown in the left panel of Fig.~\ref{fig:scheme_setup}.
We assume spin direction is isotropically distributed, while spin
magnitude is uniform in $\chi$.  Since the surrogate
model has a constraint on the mass ratio and spin
magnitudes, we are constrained in our initial mass and spin
distributions.

\figPDFmuchi

\figevolhistmassspin

Once we choose initial distribution functions, we randomly sample
$2\Nsamp$ BHs, grouped into $\Nsamp$ binaries, as depicted 
in the right panel of Fig.~\ref{fig:scheme_setup}. After the coalescences, 
these samples generate a new population (colored in
green) containing $\Nsamp$ first-generation black holes with remnant
spins $\chi_{\text{fin}}$ and masses $M_{\text{fin}}$ provided by the
surrogate code.  We keep the resulting mass-spin correlation, which
propagates to later generations, by taking the $\Nsamp$ samples of
$(M_{\text{fin}}, \chi_{\text{fin}})$ for each remnant to be the next
generation.  We however discard the remnant spin direction.  The next
step is to randomly resample (possibly with repetition) another
$2\Nsamp$ black holes into pairs, which will participate in the next
generation of mergers.  We assume isotropy and thus randomize the spin
direction of each black hole sample.  These pairs are then merged,
allowing the scheme to iterate for as many generations as desired.

Let us mention that recent research~\cite{LIGOScientific:2018jsj,
  Gerosa:2019zmo, Fishbach:2019bbm} considers selective pairings (instead of treating
each BH as independent), determined by a joint probability distribution
$p(M_{1},M_{2})$.  Present constraints on selective pairing are
consistent with each mass being selected
independently~\cite{LIGOScientific:2018jsj}, so we do not include this
in our models, leaving it as a topic for future study.
We must also add the caveat that astrophysically, BH binaries reside
in stellar clusters or galaxies, and the GW kick at merger can unbind
the remnant from the host.  Such a kicked remnant would not make it
into the next generation of merging BHs.  We ignore this effect in our
gedankenexperiment, because we are only interested in the invariant
distribution determined by the merger dynamics, unaffected by
astrophysical environments.
Further, we do not model the more realistic scenario where BHs from
multiple generations coexist in a common environment.  We only model
the transition from one generation to the next, using the generation
index as a ``time'' variable, to understand the flow of the probability 
distribution functions.

\figevolfixedmassspin

\figconvinitcond

\section{Fixed-point distributions}\label{sec:fixed-point}
We now show the evolution of populations with isotropic spins and
different initial mass distributions converging to a fixed-point distribution.
We sampled from a Gaussian, a log-normal, and a power-law distribution
to simulate a population of
$\Nsamp=$~2000 black holes.  The fixed point is a joint distribution in
spin magnitude and relative mass ratio.\footnote{The distribution of kick velocities
has low correlation with spin magnitude and relative mass ratio.
In Appendix~\ref{app:3d_distribs}, we show the independence of the kick velocity 
and the convergence of the joint three-dimensional distribution.}
We define the relative mass
ratio $\mu_{\mathrm{rel}}$ as
\begin{equation}
  \mu_{\mathrm{rel}}\equiv \frac{M_{\mathrm{fin}}}{M_{\mathrm{ini}}}
  \,,
\label{eq:mrel_def}
\end{equation}
where $M_{\mathrm{fin}}$ is the mass of the black hole remnant
and $M_{\mathrm{ini}}\equiv M_{1}+M_{2}$ is the total mass of the
progenitors.  The fraction $1-\mu_{\mathrm{rel}}$ gives
the ratio of emitted energy compared to the total initial mass of the
binary.

In the case of the power-law distribution, we used $p(m) \propto
m^{-2.3}$ corresponding to the initial Kroupa mass
function~\cite{Kroupa:2000iv, Gerosa:2017kvu, Gerosa:2019zmo,
  Gerosa:2020bjb}.
In this scenario, we restrict the mass interval to
$M\in[8;48] M_{\odot}$. This restriction is necessary since
the applicability of
\texttt{NRSur7dq4Remnant} has only been validated up to $1 \leq q\leq
6$.

\figlogevol

\figKLmassspin

In Fig.~\ref{fig:PDF_mu_chi} we show convergence to the attractive
fixed-point distribution, starting from the Kroupa initial mass
distribution and uniform in spin on the interval
$\chi\sim \mathcal{U}[0.0;0.8]$.  Solid contours contain 30\%, 68\%,
and 95\% of the probability mass, overplotted with dashed contours
from the last generation.  We can see that it takes five or six
generations to converge.  Now we would like to quantify the
convergence.  We use the Kullback-Leibler (KL) divergence~\cite{ref1}
to measure how similar are the distributions at different generations,
without marginalizing away any correlations.

In Fig.~\ref{fig:evol_hist_mass_spin}, we show the evolution of the marginalized
mass and spin magnitude distributions, from the same initial
conditions.  In the left panel, we see the mass distribution shifts to
the right as black hole masses increase through mergers.
Although the standard deviation of the mass
distribution grows, the ratio between the largest and the smallest
mass in the population tends to decrease.
In the right panel, it is hard to distinguish between the probability densities $p(\chi)$ from the third generation onwards, evidence supporting the
convergence to a fixed-point distribution.  That the spin magnitude
distribution peaks around $\chi\approx 0.7$ has also been
observed before in~\cite{Gerosa:2017kvu, Fishbach:2017dwv}, using
approximate analytical models (and in an astrophysical setting).
This even goes all the way back to the first
successful BBH merger simulations~\cite{Pretorius:2005gq}.
This is essentially a conversion of the orbital angular momentum of the two
black holes, at the time of plunge, to the spin angular
momentum of the remnant.

Convergence is more easily seen and can be tested robustly using
cumulative distribution functions (CDFs).
In Fig.~\ref{fig:evol_fixed_mass_spin}, we depict in the left panel
the evolution of the first six generations of the
$\mu_{\mathrm{rel}}$ CDF.  It immediately follows
that the fraction of energy emitted also converges to a fixed-point
distribution.  We show the spin magnitude CDF in the right panel. 

The distribution of the total BH mass tends to get
narrower (in log space) as the number of generations increases, as it loses its
memory of the initial distribution.  Most changes in the shape of the
mass and spin distributions can be quantified more precisely by
measuring moments, as reported in Appendix~\ref{app:moments}.

The Kullback-Leibler divergence, or relative entropy, is given by~\cite{ref1}
\begin{align}
  \DKL(p||q) \equiv\displaystyle{\int}p(x)\log_2\frac{p(x)}{q(x)}dx
  \,,
\label{eq:DKL}
\end{align}
and is a measure of similarity between two probability distributions
$p(x)$ and $q(x)$.
It can be interpreted as the bits of information learned when updating
prior $q(x)$ to posterior $p(x)$.
In the case we study, the parameter space is two-dimensional and is made
from the pairs $x=(\mu_{\mathrm{rel}},\chi)$. 
To estimate the integral above, we used a Gaussian kernel density
estimator (KDE) implemented in the \texttt{scipy.stats}
package~\cite{2020SciPy-NMeth}, and to minimize KDE noise, we used
a population of size $\Nsamp=40,000$.

We computed $\DKL$ for joint PDFs from (a) adjacent
generations, $\DKL(i+1 || i)$, and (b) the $i$th and the last (19th)
generation, $\DKL(19||i)$. We plot these values in
Fig.~\ref{fig:KL_mass_spin}, using the simulations initialized with a
Gaussian mass distribution in the left panel.  In both the adjacent
(case a, red) and last (case b, blue) curves, there are only two points visible in the dashed
rectangle before converging to a small value of $\DKL$,
approximately $10^{-3}$ bits of relative entropy.  We used bootstrapping to check that
this value is consistent
with sampling from a single underlying distribution. Thus, we only need
two or three iterations for the distributions to become consistent with the 
fixed-point. 

\figthermal

In the right panel of Fig.~\ref{fig:KL_mass_spin}, we use the Kroupa
power-law distribution as an initial configuration (individual
PDFs and CDFs were in Figs.~\ref{fig:evol_hist_mass_spin} and
\ref{fig:evol_fixed_mass_spin}).
Here we see the first seven generations in the dashed region in the
left necessary to converge to a fixed value of $D_{\mathrm{KL}}$, 
consistent with a gain of almost zero bits for the two forms of 
comparison shown in the red and blue curves.
Convergence is essentially equivalent if using another late generation
(excluding those in the dashed regions in
Fig.~\ref{fig:KL_mass_spin}) as a representative of the fixed-point
distribution instead of the 19th generation.

All three choices of initial mass distributions (Gaussian, log-normal,
and Kroupa) converge to the same fixed point, as seen in
Fig.~\ref{fig:conv_init_cond}.  Similarly, other initial choices of
$p(\chi)$ converge to the same fixed point distribution.  The only
difference, as seen in Fig.~\ref{fig:KL_mass_spin}, is how many
generations it takes to converge.  As seen in the left panel of
Fig.~\ref{fig:evol_hist_mass_spin}, the mass distribution approaches a
centrally-peaked distribution.  Therefore the initial Gaussian and
log-normal converge more quickly than the Kroupa power law.
Starting with a log-normal distribution shows an interesting feature,
which can be seen in Fig.~\ref{fig:log_evol}.  Each generation is
close to being log-normally distributed.  That is, merging
log-normally distributed masses generates remnants with masses that are
also drawn from a (different) log-normal distribution.

\section{Thermodynamics of distributions}\label{sec:thermal}
Now we turn to the entropy production and thermal properties (or lack
thereof) of these distributions, using the same setup as described in
Sec.~\ref{sec:setup}.  
We are interested in the fundamental
Bekenstein-Hawking properties~\cite{Bekenstein:1973ur,
  Hawking:1974sw}, rather than e.g.\ the kinetic temperature due to
motion in a cluster or galaxy. Thus we use the formulas for the Kerr
black hole,
\begin{align}
\label{eq:temp}
\frac{T_{\mathrm{BH}}}{T_{H,\odot}}&=\frac{M_{\odot}}{4\pi M}\frac{\sqrt{1-\chi^2}}{1+\sqrt{1-\chi^2}}\,, \\
\label{eq:entropy}
\frac{S_{\mathrm{BH}}}{S_{H,\odot}}&= \frac{\pi M^2}{M_{\odot}^2}\left(1+\chi^2+\sqrt{1-\chi^2}\right)\,,
\end{align}
which are in terms of the Bekenstein-Hawking temperature and entropy
scaled by the quantities $T_{H,\odot}\equiv \hbar
c^3/2Gk_{\mathrm{B}}M_{\odot}$ and  $S_{H,\odot}\equiv 4G
M^2_{\odot}/\hbar c$.

With these thermodynamic relations, we can sample black holes from
various underlying distributions.  Bianchi et~al.~\cite{Bianchi:2018ula}
considered sampling from a microcanonical ensemble, though
early-universe fluctuations would arguably draw $(M,\chi)$ from a
canonical ensemble (i.e.\ at fixed temperature) instead.  This is
possible, but the induced distribution on mass ratio has substantial
weight outside of the range $1\le q \le 6$ that we need to apply the
surrogate.  We leave this for future work.  Instead, we fix the same
uniform spin distribution as before, and use Eqs.~\eqref{eq:temp} and
\eqref{eq:entropy} as constraints for sampling at fixed temperature or
entropy.  With this approach, all of our initial configurations
satisfy the mass-ratio constraint.  We emphasize that none of our
results change qualitatively if we select other ways to distribute
spin magnitude.
In Fig.~\ref{fig:thermal}, we show the evolution of the
isothermal/isentropic states by plotting all the distributions in a
mass/spin phase diagram, analogous to a pressure/volume diagram in
ideal gas thermodynamics.  In the left panel, we show the evolution of
an initially isothermal configuration built from a uniformly
distributed set of spin magnitudes.  Isothermal states immediately
evolve into non-thermal distributions, which do not follow constant
temperature curves.
In agreement with Eqs.~\eqref{eq:temp} and~\eqref{eq:entropy}, the
temperature decreases as every black hole in the population becomes
more massive.  The same departure from thermality occurs with an
initially isentropic state depicted in the right panel of this
figure.  Similarly entropy increases, since the remnants have
increased area relative to the initial BHs. 
Departure from thermality is due to the highly non-linear dynamics of
the merger process.  A merger is irreversible, due to the area
increase theorem, so entropy is generated.  Furthermore a merger is
not quasistatic, so horizon temperature need not even be well-defined
throughout.

\figentropy

We can compute the entropy production simply via
$\Delta S = S_{\text{fin}}-S_{\text{ini}}$.  Notice that the
dimensionless entropy production depends on mass ratio as
\begin{equation}
  \label{eq:rel_ent_q}
  \frac{\Delta S}{M_{\mathrm{ini}}^2}\sim \frac{q}{(1+q)^2}
  \,,
\end{equation}
where we ignored spin factors of order 1 and dimensionful constants
like $k_{B}$.  This shows that the entropy production peaks around
$q\sim 1$ and is suppressed at extreme mass ratios.  Therefore, our
use of the surrogate model (with mass-ratio restricted to
$1\leq q\leq 6$) does not degrade our calculation of entropy
production.

We show the evolution of the entropy production rate per merger in
Fig.~\ref{fig:ent_evol}, for different choices of the initial mass
distributions.
It is interesting to observe that the growth is slower when we use the
Kroupa power-law distribution.  This distribution takes up to five or
six generations to converge, about the same number of generations as
seen in Sec.~\ref{sec:fixed-point} for convergence to the
fixed-point distributions $p(\mu_{\mathrm{rel}},\chi)$.  For each
initial condition, we see that the entropy production grows until it
reaches an approximately constant value.
The isothermal, isentropic, log-normal, and Gaussian initial conditions all behave similarly.
In all the cases we studied, once the system has converged to a fixed value for the entropy
production, it seems to have lost any memory of the initial conditions.

Black holes are by far the most entropic systems in the universe~\cite{Carroll:2004pn,
Egan:2009yy}, and thus BBH mergers are the largest contributors to the entropy
increase of the universe.  The entropy production rate in
Fig.~\ref{fig:ent_evol} allows us to estimate the entropy production
in the universe at redshift $z$, based on the BBH merger rate.
Let $\mathcal{R}$ be the differential BBH merger rate, which depends
on redshift and in principle on intrinsic parameters
$\zeta\equiv(M_{\text{ini}}, q, \vec{\chi}_{1},\vec{\chi}_{2})$ (in
the notation of Ref.~\cite{LIGOScientific:2018jsj}),
\begin{align}
  \mathcal{R}(z,\zeta)=\frac{dN}{dt\, dV\, d\zeta}
  \,.
\end{align}
Then in principle, the entropy production rate due to BBH mergers is
\begin{align}
  \frac{dS}{dt}(z) =
  \int \Delta S(\zeta) \mathcal{R}(z,\zeta)  \, dV \, d\zeta\,.
\end{align}
We have argued in Eq.~\eqref{eq:rel_ent_q} that $\Delta S$ peaks at
mass ratios close to 1, as does $\mathcal{R}$, based on astronomical
observations.  Thus as an approximation, using the value from
Fig.~\ref{fig:ent_evol}, the entropy production rate is
\begin{align}
  \label{eq:ent_gen}
  \frac{dS}{dt} \approx
  \int 1.35\pi S_{H,\odot}\left(\frac{M_{\mathrm{ini}}}{M_{\odot}}\right)^2 R(z,M_{\text{ini}})
  \, dV dM_{\text{ini}} \,,
\end{align}
where $R = dN/(dt\,dV\,dM_{\text{ini}})$, we have taken the average
over spins, and assumed the integral is dominated around $q\approx 1$.
The integrated and mass-differential merger rate has been inferred for
stellar mass BHs from Advanced LIGO and Virgo
observations~\cite{LIGOScientific:2018jsj}.  The SMBH merger rate is
currently unknown, but can be estimated from cosmological
simulations~\cite{Rodriguez-Gomez:2015aua}.  Future PTA observations
may infer this rate, and currently provide upper
limits~\cite{Arzoumanian:2018saf, Chen:2018znx}.  These limits
can not yet say if the entropy production is dominated by stellar mass
or supermassive BBHs.

\section{Discussion}\label{sec:discussion}

In this paper, we conducted a gedankenexperiment to study the
evolution of black hole distributions, due only to the intrinsic
non-linear dynamics of binary mergers.
Our main finding is the existence of an attractive fixed-point
distribution in the space of spin magnitude $\chi$ and relative mass
ratio $\mu_{\text{rel}}$ (or fraction of mass retained).
We used the surrogate routines in \texttt{NRSur7dq4Remnant} to compute
the remnant mass and spin of each merger in the population after every
iteration/generation.
Although we used stellar-mass scale black holes as examples in this
work, the thought experiment itself is scale-free: it can extend up to
supermassive black holes or down to primordial ones.
Our interest is in the underlying fixed-point distribution due to
merger dynamics, so we did not attempt to capture real astrophysical
effects like unbinding from a cluster or galactic halo.

All initial distributions we studied (Gaussian, log-normal, Kroupa,
isothermal, isentropic) converged to the same fixed point
distribution.  The only difference was the number of generations
needed for convergence.  We quantified this convergence with the
Kullback-Leibler divergence.

We studied the (lack of) thermality of the fixed-point
distribution, from the point of view of black hole thermodynamics.
Starting with states of constant Bekenstein-Hawking temperature or
entropy, after only one generation the resulting distribution is
neither isothermal nor isentropic.  This is a result of the
non-linear merger dynamics being neither quasi-static nor reversible:
a merger is not a thermal process.  We computed the average entropy
production per merger in these distributions -- which, as an aside,
are likely the largest sources of entropy in the universe.  We found
that the average entropy production per merger (in units of the
initial mass squared) converged to a constant, again independent of
initial conditions.
This can be used to approximate the rate of entropy production of the
universe.

There are several possibilities present for future studies extending
this work.  First, we assumed that each black hole is drawn
independently, rather than the more sophisticated possibility of
selective pairing.  Selective pairing might lead to a different
fixed-point.  Second, in constructing our isothermal states, we did not sample
from a canonical ensemble, which would be the natural approach.  This
was due to the surrogate's mass-ratio limitation $1\le q \le 6$.  An
improved surrogate would make it possible to test the canonical
ensemble.  Third, in all of our simulations, we assigned initial conditions
at the surrogate's reference time.  These can be connected to initial
conditions at infinite separation using the recent work
of~\cite{Reali:2020vkf}, though we expect the fixed point to remain
the same with this change of initial conditions.  Finally,
Ref.~\cite{Christian:2018mjv} presents a different approach to compute
the evolution of mass distributions, following a coagulation
equation. This procedure provides an alternative way to corroborate
our results.

\acknowledgments
We would like to thank
Emanuele Berti,
Eugenio Bianchi,
Luca Bombelli,
Andrei Frolov,
Davide Gerosa,
Anuradha Gupta,
Levon Pogosian,
Vijay Varma,
Alex Zucca,
and a number of anonymous referees
for their comments and suggestions on a previous draft of this paper.
This work made use of the \texttt{NumPy}~\cite{2020NumPy-Array} and
\texttt{SciPy}~\cite{2020SciPy-NMeth} packages.

\appendix

\section{3D joint distribution and independence of the kick velocity}
\label{app:3d_distribs}

\figfulljointPDF

\figvindep

Kick velocity is also an output of the surrogate code. It is,
therefore, reasonable to evaluate the fixed-point convergence of the
three-dimensional joint distribution $p(\mu_{\mathrm{rel}}, \chi, v/c)$
as plotted in Fig.~\ref{fig:corner_mu_chi_v}. The panel showing the 
marginalized distribution $p(\mu_{\mathrm{rel}},\chi)$ shows the same
convergence seen in Fig.~\ref{fig:PDF_mu_chi}, in addition to the panels 
$p(v/c,\chi)$ and $p(v/c,\mu_{\mathrm{rel}})$ which are also
consistent with the convergence to a fixed-point distribution. This corner plot
was produced using $N_{\mathrm{samp}}=40,000$ instead of the 2000
samples used throughout much of the paper. This change
did not affect any of our results. The bottom row in this corner plot
shows that the kick velocity distribution also converges to a fixed-point.

\figmoments

The marginalized distributions $p(v/c,\chi)$ and $p(v/c,\mu_{\mathrm{rel}})$ 
show low correlations between the kick velocity and either $\chi$ or $\mu_{\mathrm{rel}}$.
To verify this, we created a de-correlated distribution by starting with
the fixed-point distribution, and permuting the values of $v/c$ among
the samples.  This does not affect the marginalized distribution
$p(v/c)$, while breaking any correlations between $v/c$ and all other variables.
We show 30\%, 68\%, and 95\% contours of the original and
de-correlated distributions in Fig.~\ref{fig:v_indep}.
The two sets of contours are very similar,
suggesting the remnant kick velocity is approximately independent of
the remnant spin and relative mass ratio.

It is also possible to quantify the convergence of the three-dimensional joint distributions by
computing the KL divergence, as indicated in Section \ref{sec:fixed-point}.
We found a decrement in the magnitude of $D_{\mathrm{KL}}$ as the number of 
generations increases, which is similar to our results in Fig.~\ref{fig:KL_mass_spin} for the joint
$(\mu_{\mathrm{rel}},\chi)$ distribution. The minimum value of $D_{\mathrm{KL}}$ grows 
with the number of dimensions of parameter space.  Kernel density
estimation is also noisier in higher dimensions, so the 3D results are
noisier than the 2D results.  Still, convergence to a fixed-point happens in
the same number of generations.

\section{Evolution of moments of distributions}\label{app:moments}

Another way to quantify the convergence of these distributions is to
evaluate various central moments of the one-dimensional
marginalized distributions. With the average mass as $\bar{m} = \avg{m}$, 
we define the $k$th central mass moment as $m_{k}\equiv \avg{(m-\bar{m})^{k}}$, 
and similarly for spin. These quantities moments carry dimensions 
of mass (or spin) to the power $k$, so they can be made dimensionless 
by computing $|m_{k}|^{1/k}/\bar{m}$ (and likewise for spin).
We plot these dimensionless central moments in Fig.~\ref{fig:moments},
as a function of generation, using the Kroupa power law as the
initial condition on mass.

In the left panel, note that all of the moments decrease until a large number of generations.
In particular, the ratio of the variance with the mean mass,
$m_2/\bar{m}$, is consistent with narrowing of the marginal mass
distribution visible in Figs.~\ref{fig:evol_hist_mass_spin} and \ref{fig:log_evol}.
After twelve generations of mergers, the shape of the mass
distribution is approximately constant. The number of generations needed to observe the
approximate invariance of the mass distribution shape reduces if we
choose (log-)normal or thermal initial conditions. The scaled central moments
of the marginal spin distribution appear in the right panel, and also converge
after five generation, as seen previously. In both panels, skewness
($m_3$ and $s_3$) is not dominant, though, not all odd moments are negligible. 
Thus, even when the fixed-point distribution is centrally-peaked, it is not a Gaussian.

\bibliography{Distnotes}
\end{document}